# Acousto-optic volumetric gating for reflection-mode deep optical imaging within a scattering medium

Hakseok Ko[1,2], Junghoon Kim[1,3], Jin-Hee Hong[1], Junyeob Cheon[4], Seungwoo Lee[4,5], Mooseok Jang[2,*], Wonshik Choi[1,3,*]

[1]*Center for Molecular Spectroscopy and Dynamics, Institute for Basic Science (IBS), Seoul 02481, Republic of Korea*
[2]*Department of Bio and Brain Engineering, Korea Advanced Institute of Science and Technology (KAIST), 291, Daehak-ro, Yuseong-gu, Daejeon 34141, Republic of Korea*
[3]*Department of Physics, Korea University, Seoul 02481, Republic of Korea*
[4]*KU-KIST Graduate School of Converging Science and Technology, Korea University, Seoul 02481, Republic of Korea*
[5]*Department of Integrative Energy Engineering (college of engineering) and Department of Biomicrosystem Technology, Korea University, Seoul 02481, Republic of Korea*
[*]*e-mail: mooseok@kaist.ac.kr, wonshik@korea.ac.kr*

**Abstract**

The imaging depth of deep-tissue optical microscopy is governed by the performance of the gating operation that suppresses the multiply scattered waves obscuring the ballistic waves. Although various gating operations based on confocal, time-resolved/coherence-gated, and polarization-selective detections have proven to be effective, each has its own limitation; certain types of multiply scattered waves can bypass the gating. Here, we propose a method, volumetric gating, that introduces ultrasound focus to confocal reflectance imaging to suppress the multiply scattered waves traveling outside the ultrasonic focal volume. The volumetric gating axially rejects the multiply scattered wave traveling to a depth shallower than the object plane while suppressing the deeper penetrating portion that travels across the object plane outside the transversal extent of the ultrasonic focus of $30 \times 90$ μm$^2$. These joint gating actions along the axial and lateral directions attenuate the multiply scattered waves by a factor of 1/1000 or smaller, thereby extending the imaging depth to 12.1 times the scattering mean free path while maintaining the diffraction-limited resolution of 1.5 μm. We demonstrated an increase in the imaging depth and contrast for internal tissue imaging of mouse colon and small intestine through their outer walls. We further developed theoretical and experimental frameworks to characterise the axial distribution of light trajectories inside scattering media. The volumetric gating will serve as an important addition to deep-tissue imaging modalities and a useful tool for studying wave propagation in scattering media.

**Introduction**

Deep optical imaging with a high spatial resolution at the diffraction limit typically relies on the detection of ballistic waves traveling straight through a scattering medium for the image formation of an object embedded therein[1]. Because the ballistic wave intensities are attenuated exponentially with depth, they are obscured by multiply scattered waves even at depths shallower than a few scattering mean free paths. Confocal[2,3], temporal/coherence[4–8], and polarization[9,10] gating operations have been introduced to suppress the multiply scattered waves. These methods exploit the distinctive properties of the ballistic waves, such as momentum conservation, one-to-one relation between their penetrating depth and flight time, and polarization-state preservation, to suppress them. However, certain types of multiply scattered waves can bypass the gating operations, thereby limiting the imaging depth. For example, a fraction of the multiply scattered waves can arrive at the confocal pinhole with the same polarization state as the ballistic waves. Similarly, the multiply scattered waves traveling at depths shallower than the object plane can have the same flight time as the ballistic waves reflected at the object plane, thereby bypassing the temporal/coherence gating. In fact, the innate limitations of all these gating operations arise, because the existing gating operations take place outside the scattering medium.

The introduction of ultrasound to optical imaging has gained attention over the past decades as a suitable method to extend the depth limit of optical imaging. Ultrasound can penetrate deeper than light such that an acoustic focus can be created at the depths of 1–10 cm inside the scattering medium. The selective detection of the acoustically modulated optical waves enables the gating of optical waves within the scattering medium. Therefore, this method can enhance the depth of the optical imaging and focusing up to that of the ultrasound waves, while their spatial resolution is intrinsically set by the acoustic focus[11–13]. To date, many approaches have been proposed based on linear algebra operations to gain optical-scale resolution while retaining the acoustic imaging depth[14–16]. In our earlier study, we proposed a lateral gating method based on the acousto-optic modulation, in which the acousto-optically modulated ballistic waves were selectively detected to suppress the effect of multiply scattered waves in optically diffraction-limited imaging. However, all these studies have been conducted in transmission-mode imaging, confining their applicability to tissue sections and small animals such as zebrafish. Their extension to reflection-mode imaging will facilitate *in vivo* studies, where optical detectors cannot be placed on the opposite side of the illumination. However, it is much more challenging to detect the acousto-optically modulated ballistic waves in the reflection mode because the optical back-reflection signal is weaker than the transmission signal in deep-tissue imaging, which is exacerbated by the low ultrasound modulation efficiency (~1 %).

Here, we propose volumetric gating microscopy that integrates ultrasound-modulated optical detection into confocal reflectance microscopy to suppress the multiply scattered waves traveling outside the volume of the ultrasound focus (30($x$) × 90($y$) × 30($z$) μm$^3$) confined in a three-dimensional (3D) space. The volumetric gating removes all the multiply scattered waves traveling to a depth shallower than the object plane ($x$-$y$ plane)—a unique feature appearing only in the reflection-mode imaging. Furthermore, it also suppresses a substantial fraction of the multiply scattered waves reaching the object plane or deeper by the lateral area ratio between the lateral ultrasound focus size (30($x$) × 90($y$) μm$^2$) to the lateral spread of the multiply scattered wave. The combined action of the axial and lateral gating makes the volumetric gating more effective at deeper depths. It enabled us to enhance the ballistic signal to background noise (i.e. multiply scattered waves) ratio (SBR) by more than a factor of 1,000, making it possible to obtain optically diffraction-limited reflectance imaging up to the optical depth of 12.1 times the scattering mean free path. We proved the increase in imaging depth and image contrast by imaging crypts in the colon and villi in the small intestine through their outer walls. Notably, we could quantify the traveling wave distribution within a scattering medium by combining the volumetric gating in reflection mode and lateral gating in transmission mode. The proposed volumetric gating is an important addition to deep optical imaging, and the experimental framework developed in our study provides a new tool for interrogating the wave propagation within a scattering medium.

## Results

**Principle of volumetric gating in confocal reflectance imaging**

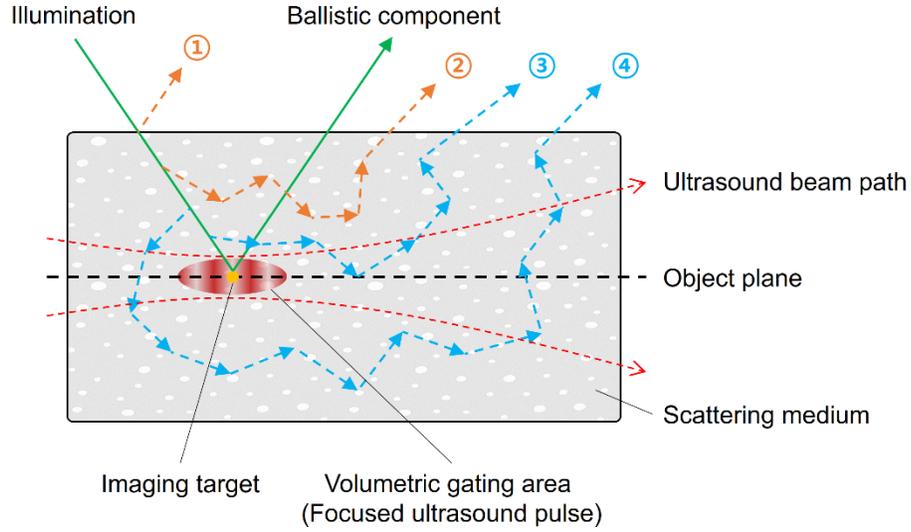

**Figure 1. Principle of volumetric gating.** Multiply scattered waves from a scattering medium can be classified into four categories by the maximum depth at which the light component reaches: ① specular reflection from the incident surface, ② reflection at depths shallower than the object plane, ③ reflection at the object plane, and ④ reflection at depths deeper than the object plane. Volumetric gating completely rejects ① and ②. It further gates out the fraction of ③ and ④ traveling outside the acoustic focus. The gating efficiency depends on the ratio between the area of the diffused light and the lateral cross-section of the acoustic focus at the object plane.

Volumetric gating picks up the confocal reflectance signal modulated by the ultrasound focus introduced in the object plane. In confocal reflectance imaging in scattering media, the associated multiply scattered waves can be categorised into four types depending on their maximum traveling depths (as shown in Fig. 1): ① specular reflection at the surface, ② traveling depths shallower than the object plane, ③ traveling depth to the object plane, and ④ traveling depths larger than the object plane. ① and ② are completely rejected by the volumetric gating, while ③ and ④ are partially rejected by the ratio of the transverse area of the acoustic focus $A_{AF}$ at the focal plane and the lateral area $A_{MS}$ of the corresponding multiply scattered wave components (i.e. the factor of $A_{AF}/A_{MS}$). With the volumetric gating operation, only the ballistic component (green arrow) and the fraction of ③ and ④ traveling inside the acoustic focus are detected and contribute to the signal and the noise in image formation, respectively.

The effect of volumetric gating can be characterised by the ballistic signal $I_B$ to the multiply scattered wave background ratio (SBR). Setting $I_S$ as the combined intensity of ① and ② from the shallower depths and $I_D$ as the intensity of ③ and ④ from the deeper depths, the SBRs without and with the volumetric gating are given by $\tau = \frac{I_B}{(I_S+I_D)}$, and $\tau_{VG} = \frac{A_{MS}}{A_{AF}} \frac{I_B}{I_D}$, respectively. The factor representing the SBR enhancement by the volumetric gating is then given by

$$\eta_{VG} = \frac{\tau_{VG}}{\tau} = \frac{A_{MS}}{A_{AF}}(1 + I_S/I_D) = \eta_T(1 + I_S/I_D). \quad (1)$$

Here $\eta_T = A_{MS}/A_{AF}$ is the SBR enhancement of the lateral gating operation in transmission-mode imaging[17]. Because all the detected waves in the transmission geometry travel through the scattering medium, there is no action of rejecting multiply scattered waves traveling at depths shallower than the object plane. Therefore, $\eta_{VG}$ is always larger than $\eta_T$. The efficiency of the SBR enhancement is set by $I_S/I_D$ as well as $\eta_T$. It is noteworthy that $I_S/I_D$ tends to increase as the target imaging depth is increased. This suggests that the benefit of volumetric gating is more pronounced with the increase in imaging depth. As evident from our experiments, we could independently measure $\eta_{VG}$ and $\eta_T$ from the volumetric gating in the reflection mode and lateral gating in the transmission mode, from which $I_S/I_D$ could be quantified depending on the depth of the ultrasound focus. Here, $I_S/I_D$ is the ratio of the intensity of the multiply scattered waves traveling at depths shallower than the object plane to that of the waves traveling at depths deeper than the object plane. This intensity ratio directly indicates the internal axial distribution of the back-scattered optical waves.

It is worthwhile to compare the mechanisms of volumetric gating and temporal gating. In time gating (or coherence gating), the multiply scattered wave components ① and ④ can be fully rejected while a fraction of ② and ③ remain unfiltered. Multiply scattered waves from shallower depths can have the same flight time as the ballistic signal wave. Furthermore, a finite time gating window allows the detection of the multiply scattered waves reaching the object plane. In other words, the performance of time gating is degraded as the fraction of multiply scattered waves traveling to depths shallower than the object plane increases[18]. This is the clear distinction between time gating and volumetric gating, suggesting that volumetric gating can be a formidable addition to deep imaging modalities, where temporal gating is compromised.

**Experimental setup of volumetric gating microscopy**

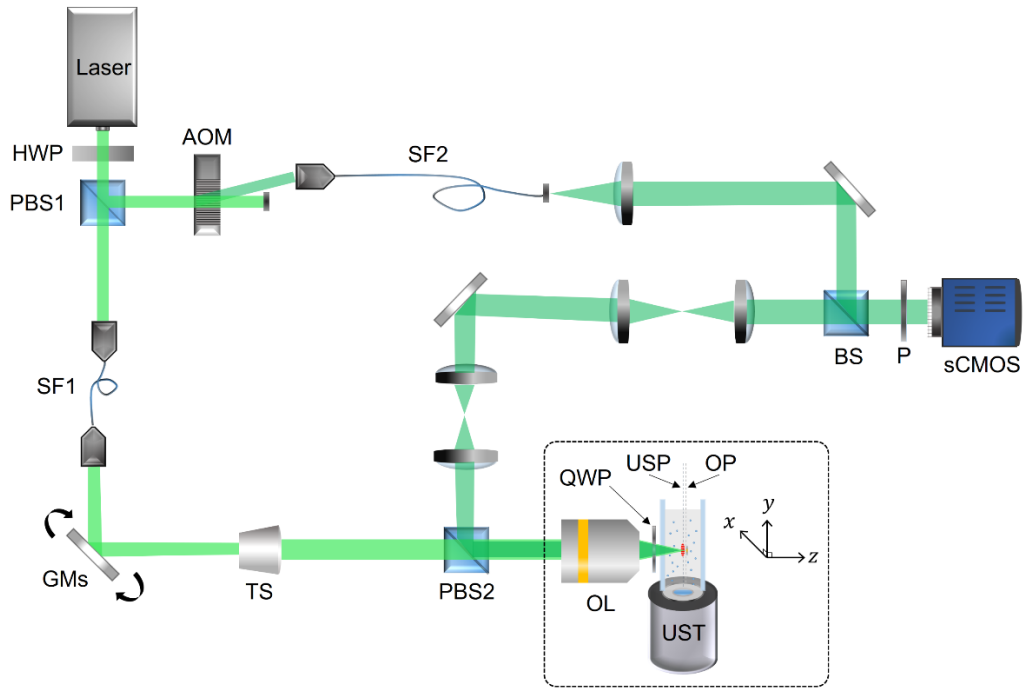

**Figure 2. Experimental setup of volumetric gating microscopy.** The experimental setup is implemented by introducing an ultrasonic transducer on a confocal reflectance interferometric microscope. To implement confocal reflectance imaging, the camera was placed at the plane conjugate to the object plane, and the interference intensity at the camera pixels conjugate to the focused illumination spot was taken as the image signal. In a Mach–Zehnder interferometer, both the reference and sample beams are spatially filtered through a single mode fibre. In the reference beam path, an acousto-optic modulator is placed to induce selective interference with the acoustically modulated signal from the sample beam path. HWP: half-wave plate, QWP: quarter-wave plate, PBS1, 2: polarizing beam splitters, BS: beam splitters, AOM: acousto-optic modulator, SF1, 2: spatial filters via single-mode optical fibres, GMs: 2-axis galvanometer scanning mirror, TS: telescope, OL: objective lens, UST: ultrasound transducer, USP: ultrasound propagation plane, OP: object plane set by the objective lens focal plane, P: linear polariser, and sCMOS: camera.

Our optical system is a reflectance-mode interferometric laser scanning microscope designed for selective detection of ultrasound-modulated backscattering signals (Fig. 2). We used a 532 nm pulse laser (Edgewave BX60-2-G, pulse width: 7 ns) as the light source. The output beam from the laser was divided into sample and reference beams at a polarizing beam splitter (PBS1). A half-wave plate was used to tune the powers of the sample and reference beams, in conjunction with the PBS1. The sample beam was spatially filtered at the spatial filter (SF1) via coupling light to a single-mode optical fibre. The output beam from the optical fibre was expanded through the telescope lens pair. A two-axis galvanometer scanning mirror (GM, Thorlabs GVS-001) controlled the incidence angle of the sample beam to scan a focused illumination at an object plane (indicated by a dashed line in Fig. 2) through an objective lens (Mitutoyo M

Plan Apo 10X NA 0.28). Due to the limited physical aperture geometrically constrained by the transducer, the numerical aperture was reduced to 0.18, setting the diffraction-limited spatial resolution to 1.5 μm. The backscattered wave from the sample was relayed to a camera (sCMOS, PCO edge 4.2 CLHS) placed at a plane conjugate to the object plane. The magnification from the object plane to the camera was 17.1. A quarter-wave plate was placed between the objective lens and sample to prevent stray reflections from the optical elements, except those from the sample, in conjunction with PBS2 (a second PBS).

A focused ultrasonic wave was introduced at the object plane to modulate the sample beam traveling through a small volume. Specifically, an ultrasound transducer (UST, Olympus V3330) focused three cycles of the acoustic pulses propagating perpendicular to the sample beam. The focal volume of the ultrasound was $30(x) \times 90(y) \times 30(z)$ μm$^3$. The ultrasound propagation plane was matched with the object plane for the biological samples, but it was placed slightly upstream towards the objective lens in the case of the rigid resolution target sample to avoid impedance mismatch. The sample beam (i.e. the optical pulse) and the acoustic pulse were synchronised by a master clock such that they coincided at the sample position. For the selective detection of the ultrasound-modulated sample waves, we prepared a reference beam, whose frequency was shifted by a frequency (50.01 MHz) identical to that of the UST using an acousto-optic modulator. The reference beam was spatially filtered at SF2 via coupling to a single-mode optical fibre and relayed to the camera to form an interference image with the sample beam. Phase-shifting holography was implemented to obtain the phase and amplitude of the ultrasound-modulated optical waves by recording four interference images for the reference beam phase shifts of $\Delta\phi = 0, \pi/2, \pi,$ and $3\pi/2$[19,20]. For confocal volumetric gating imaging, the signal at the camera pixels conjugate to the focused illumination spot at the object plane was selected for image reconstruction.

**Demonstration of the volumetric gating imaging**

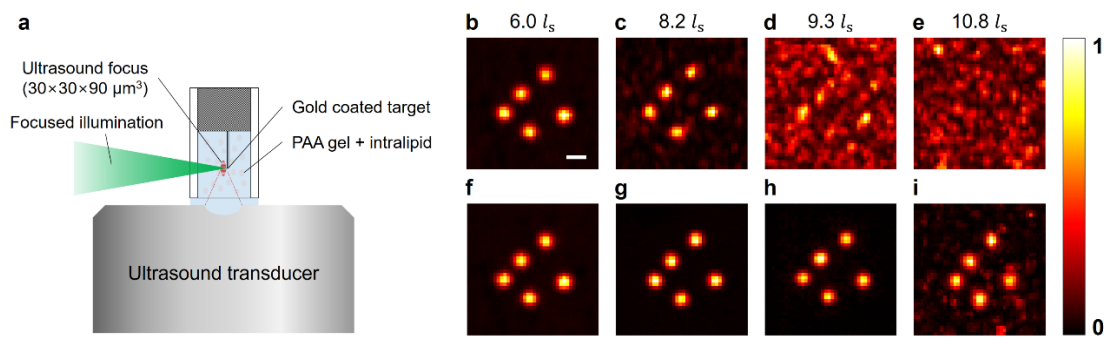

**Figure 3. Demonstration of imaging inside thick scattering media with volumetric gating. a**, Schematic of the sample arrangement, consisting of gold-coated reflective targets embedded within a thick scattering

medium at a fixed depth of 4 mm. Ultrasound focus was placed ~ 10 μm in front of the imaging targets to avoid ultrasonic attenuation due to acoustic impedance mismatch. Each image was reconstructed by scanning 1600 points within a 20 × 20 μm² FOV. **b-e,** Reconstructed intensity images of five 2-μm-diameter gold-coated circles obtained without volumetric gating. The optical thicknesses of the scattering medium in front of the object plane were 6.0, 8.2, 9.3, and 10.8 $l_s$, respectively. **f-i,** Reconstructed images with the volumetric gating in the same configurations as in **b-e**. Each image was normalised by its maximum intensity. Scale bar: 3 μm.

To demonstrate the effect of volumetric gating in suppressing the multiply scattered waves, we conducted confocal reflectance imaging of gold-coated targets embedded in a scattering medium. The scattering medium was made of a polyacrylamide (PAA) gel mixed with fat emulsion (Intralipid). The concentration of the fat emulsion in the PAA gel was varied to control the optical thickness of the medium (see Methods for the sample preparation). The reflective imaging target consisted of five gold-coated circles patterned on a slide glass. The diameter of each circle was 2 μm, and the average reflectance of the gold layer was 80 %. The target was placed at a focal plane of the objective lens located at a depth of 4 mm within a scattering medium (Fig. 3a). We placed the ultrasound focus slightly upstream towards the objective lens to prevent the slide glasses from obstructing the ultrasound focus because of the impedance mismatch between the PAA gel and the slide glass. The distance from the object plane to the centre of the acoustic focus was set to less than 10 μm, which is the minimum distance free from obstruction.

Figures 3b–e show the interferometric confocal reflectance images obtained without volumetric gating. In this case, the image contrast was significantly decreased at the depth of 8.2 $l_s$, and the target object appeared to be completely obscured by the multiply scattered waves at increased optical depths. Figures 3f–i show the confocal reflectance imaging with volumetric gating. In this case, the image contrast at the depth of 8.2 $l_s$ was three times higher than that obtained without volumetric gating, and the circular targets were clearly visible up to a depth of 10.8 $l_s$ with an image contrast of 22. These results support the advantage of volumetric gating in enhancing the imaging depth within scattering media by attenuating the multiply scattered waves. The achievable imaging depth was 11.7 $l_s$, which is slightly higher than the previous result of 11.5 $l_s$ obtained via time gating[21].

**Quantitative assessment of the multiple scattering suppression**

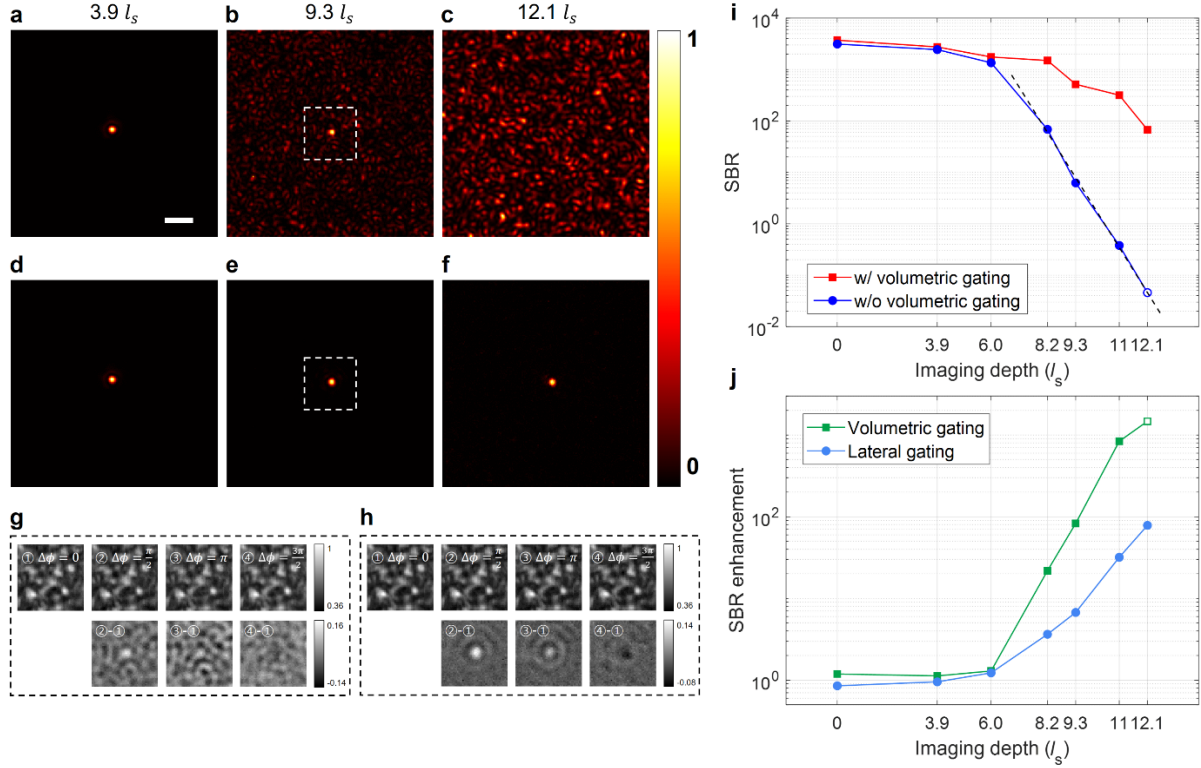

**Figure 4. Quantitative assessment of multiple scattering suppression by volumetric gating. a-c,** PSFs measured without volumetric gating when the optical thicknesses were 3.9, 9.3, 12.1 $l_s$, respectively. **d-f,** PSFs with volumetric gating for the same configurations as **a-c.** Each image was normalised to its maximum intensity. Scale bar: 10 μm. **g,** Raw interference images for reconstructing the complex field map in a dashed rectangular box in **b**. Four images were captured using phase-shifting interferometry with the phase of the reference beam shifted by $\Delta\phi=$ 0, $\pi/2$, $\pi$, and $3\pi/2$. The differences in the interference images with respect to that taken at $\Delta\phi=0$ are shown in the bottom row. **h,** Same as **g**, but for reconstructing the complex field map in a dashed rectangular box in **e** with the volumetric gating. **i,** Ratio of the ballistic signal intensity to the multiply scattered waves (SBR) versus the imaging depth with volumetric gating ($\tau_{VG}$, red squares) and that without volumetric gating ($\tau$, blue circles). The blue open circle indicates the $\tau$ at 12.1 $l_s$ estimated by assuming an exponential attenuation of the ballistic intensity (black dashed line). **j,** SBR enhancement factor of the volumetric gating, obtained from $\eta_{VG} = \tau_{VG}/\tau$ in **i** (Green squares). The green open square represents the value estimated from the blue open circle in **i**. The blue circles indicate the lateral noise suppression factor $\eta_T$ measured by the transmission-mode gating.

We quantitatively assessed the multiple scattering suppression of the volumetric gating at different imaging depths by measuring the point-spread function (PSF) using a focused illumination to a plane mirror at the OP. Figures 4a–f show the PSFs measured at the camera plane without (Figs. 4a–c) and with (Figs. 4d–f) the volumetric gating for the depths of the object plane of $3.9l_s$, $9.3l_s$, and $12.1l_s$. Without volumetric gating (Figs. 4a–c), the PSFs were degraded rapidly as the imaging depth increases. However, with the volumetric gating (Figs. 4d–f), the diffraction-limited PSF was maintained even at the depth of 12.1 $l_s$ owing to the suppression of the multiply scattered waves.

To elucidate the effect of volumetric gating, we obtained raw interference images as shown in Figs. 4g and h that were used for obtaining Figs. 4b and e, respectively (see Method for detailed procedure of data acquisition for holographic images). The first row shows the four interference images acquired at the phase shift $\Delta\phi$ of the reference beam in the integer multiples of $\pi/2$. The second row shows the differential images of the phase-shifted images relative to that taken at $\Delta\phi=0$. It is noteworthy that the raw interference images appeared the same, irrespective of the volumetric gating. This is because there were intense multiply scattered waves whose optical path lengths relative to the reference beam path were outside the coherence gating window of ~1.5 mm set by the coherence length of the light source. However, the differential images could be clearly distinguished. Without the volumetric gating (Fig. 4g), the distinctive intensity fluctuations (i.e. speckle patterns) of the multiply scattered waves were mainly visible in the differential images. With the volumetric gating (Fig. 4h), the interference patterns of the ballistic signal were clearly visible in the differential images, supporting a substantial increase in the SBR.

Based on the measured PSFs, we quantitatively analysed the contributions of the ballistic and multiply scattered waves with and without the volumetric gating. We measured the SBR by estimating the intensity ratio between the ballistic wave and the multiply scattered wave from the peak intensity of the focused spot and the average intensity of the surrounding background speckle pattern without ($\tau$) and with ($\tau_{VG}$) the volumetric gating (Fig. 4i). To increase the measurement fidelity by ensemble averaging, hundred PSFs were measured at different illumination positions and subsequently averaged. Up to the depth of 6.0 $l_s$, $\tau$ and $\tau_{VG}$ were almost the same, because the coherence gating of the 1.5-mm window was effective in rejecting most of the multiply scattered waves. With further increase in the imaging depth, $\tau$ dropped rapidly, reached almost unity at 10.8 $l_s$, and became hardly measurable at 12.1 $l_s$. On the contrary, $\tau_{VG}$ was as high as ~100 even at 12.1 $l_s$, resulting in the diffraction-limited PSF.

To quantify the effect of volumetric gating on the multiple scattering suppression, the SBR enhancement ratio $\eta_{VG}$ was calculated as $\tau_{VG}/\tau$ and plotted in Fig. 4j (green squares). As evident, it steadily increased with the imaging depth, and at the depth of 11.0 $l_s$, there was an 840-fold enhancement. These results show that the effect of volumetric gating is more pronounced at a deeper depth. It is difficult to characterise $\eta_{VG}$ at 12.1 $l_s$ based on the measured PSFs, because the SBR without the volumetric gating, $\tau$, was below the detection limit. More specifically, the ballistic wave could not be detected owing to the presence of intense multiply scattered waves, as shown in Fig. 4c. Alternatively, the ballistic contribution can be estimated based on the exponential decay of the ballistic signal (blue open circle in Fig. 4i), resulting in the estimated

SBR enhancement of $\eta_{VG} = 1{,}446$ at 12.1 $l_s$ (green open square in Fig. 4j). The SBR enhancement ($\eta_{VG}$) in the reflection-mode imaging was significantly larger than $\eta_T$ of the lateral gating in the transmission-mode experiments (blue circles in Fig. 4j) for the same sample configurations. As discussed earlier, the lateral spread of the multiply scattered waves relative to the lateral size of the acoustic gating governs the noise suppression ratio in the transmission mode[17]. On the contrary, the multiply scattered waves from depths shallower than the acoustic focus were additionally rejected in the reflection-mode operation. According to Eq. (1), $\eta_{VG}/\eta_T = 1 + I_S/I_D$. At 11.0 $l_s$, $\eta_{VG} = 839$ and $\eta_T = 32$ such that $I_S/I_D = 25$. This indicates that the axial gating accounted for 96 % of the multiple scattering rejection. In fact, we could obtain $I_S/I_D$ depending on the imaging depth, or the position of the ultrasound focus, and observed a steady increase up to 11.0 $l_s$ and a slight reduction thereafter. This presents the first experimental observation of the internal optical field distribution within a 3D scattering medium.

**Volumetric gated imaging of biological tissues**

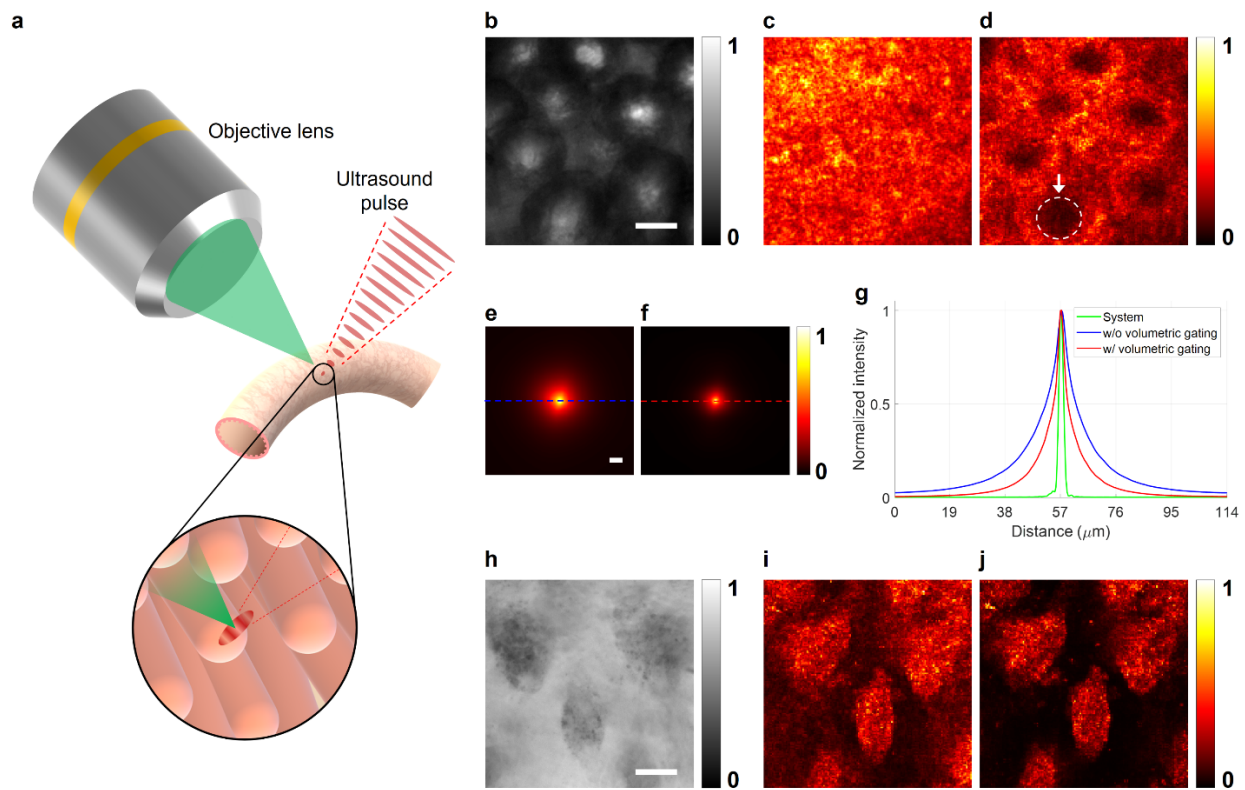

**Figure 5. Volumetric gating imaging of mucosa and villi from the outer walls of the colon and small intestine. a,** Schematic of the measurement geometry. An objective lens focused a pulsed laser on the mouse colon and intestine in the internal walls through the outer walls and collected the reflected light waves. A focused ultrasound pulse was generated at the focal point of the objective lens for the volumetric gating. We measured the fixed mouse colon and intestine. **b,** Brightfield transmission microscope image of a mouse colon, obtained by illuminating a 532-nm LED from the opposite side of the objective lens in **a**.

Scale bar: 30 μm. **c-d,** Reconstructed confocal reflectance images of a mouse colon without and with the volumetric gating, respectively. Detection pinhole size: 10 μm. **e-f,** Ensemble averaged PSFs without and with the volumetric gating, respectively. Scale bar: 10 μm. Each image was normalised by its maximum intensity. **g,** Line profiles of the PSFs in **e** and **f** along with that of the system PSF. **h,** Brightfield transmission microscope image of a mouse intestine. Scale bar: 60 μm. **i-j,** Reconstructed confocal reflectance images of a mouse intestine without and with the volumetric gating, respectively. Detection pinhole size: 10 μm.

To validate the applicability of the proposed volumetric gating method in biological imaging, we imaged the rat colon and small intestine. To demonstrate the deep-tissue imaging capability of the proposed method, we performed imaging by focusing the laser pulses from the outer walls to the mucosa and villi surfaces in the inner walls (Fig. 5a). The acoustic focus was configured to be orthogonal to the optical axis set by the objective lens. The reflected light was collected by the same objective lens and relayed to the volumetric gating detection system. The confocal reflectance images were acquired over the field of view (FOV) of $10 \times 10$ μm$^2$ in the object plane at a time, either without or with the volumetric gating. In the case of colon imaging, $15 \times 15$ images were stitched to form the images with the FOV of $150 \times 150$ μm$^2$ (Fig. 5c without volumetric gating, and Fig. 5d with volumetric gating). In the case of small intestine imaging, $30 \times 30$ images were stitched to form the images with the FOV of $300 \times 300$ μm$^2$ (Fig. 5i without volumetric gating, and Fig. 5j with volumetric gating). As a point of reference, we recorded the transmission-mode images of the colon and intestine (Figs. 5b and h, respectively) by illuminating them with a 532-nm light-emitting diode (LED) from the opposite side of the objective lens. Because there is only one-way scattering in the transmittance imaging, its imaging fidelity is higher than that of reflection-mode imaging. Therefore, transmittance imaging can serve as the ground truth for volumetric-gating-assisted reflection-mode imaging.

In the case of colon imaging, the crypts in the mucosa were clearly visible in the volumetric gating images (white arrows in Fig. 5d), while no discernible structures were observed in the images obtained without volumetric gating (Fig. 5c). The crypts appeared dark because the objective focus was set on the surface of the mucosa. The transmission-mode images presented in Fig. 5b show the opposite contrast. In the transmission-mode images, the crypts were brighter due to their smaller tissue thickness than the surrounding area. The positions of the crypts in the transmittance image matched well with those in the volumetric gating image, supporting the validity of the volumetric imaging (images shown in Fig. 5c). Because the crypts were invisible without the volumetric gating, it was not possible to estimate the enhancement of SBR from the images acquired by volumetric gating. In the small intestine images, individual villi were resolved even without the volumetric gating, because the wall thickness was smaller than that of the colon (Fig. 5i). The confocal reflectance image obtained by volumetric gating revealed an enhanced SBR by a factor of four due to the suppression of the multiply scattered wave (Fig. 5j). The transmittance image in Fig. 5h also supports the validity of the confocal reflectance images shown in Figs. 5i and j.

We analysed the effect of volumetric gating by comparing the ensemble-averaged PSFs without and with the volumetric gating as shown in Figs. 5e and f, respectively. Their line profiles are compared in Fig. 5g, where the PSF width is reduced from 10.3 to 5.0 μm after the volumetric gating. This suggests that volumetric gating effectively suppresses the multiply scattered waves. However, the measured PSFs were much broader than the diffraction-limited PSF of the system (green curve in Fig. 5g). A similar trend was observed in the small intestine image, where the PSF width was reduced from 4.7 to 3.2 μm after the volumetric gating, respectively. This is partly because the aberrations induced by the tubular shape of the colon and small intestine cause broadening of the waves that were singly scattered by the mucosa and villi. In addition, the reflectance of the surface of the mucosa and villi was very low such that the aberration-free ballistic wave was obscured by the multiply scattered waves even after the volumetric gating—this is the innate challenge in label-free reflectance imaging of biological tissues. Interestingly, the detection pinhole size played an important role in determining the image contrast in the presence of weak reflectance and aberration. We selected the detection pinhole sizes of 10 μm for obtaining both the colon and small intestine images shown in Fig. 5.

**Discussion**

We implemented confocal reflectance imaging of the ultrasound-modulated backscattered optical waves to eliminate the multiply scattered waves traveling outside the ultrasound focus (30 × 30 × 90 μm³) confined in a 3D space within the scattering medium. All the previously reported acousto-optic approaches, implemented in the transmission geometry, provided lateral gating set by the cross-sectional area of the ultrasound focus in the object plane. In contrast, our reflection-mode imaging offers a unique axial gating that eliminates the multiply scattered waves traveling at a depth shallower than the ultrasound focus. We demonstrated an increase in the ballistic signal to multiple scattering background ratio (SBR) by more than a factor of 1000 via volumetric gating, which enabled the realization of diffraction-limited optical imaging (1.5 μm spatial resolution) up to a depth of 12.1 times the scattering mean free path. We observed that the SBR enhanced with the increasing depth, and our analysis suggests that the axial gating was responsible for as large as 96 % of the total multiple scattering suppression.

Because of the reflection-mode imaging configuration, the proposed volumetric gating can be readily applied to the evaluation of thick biological tissues. In this study, we imaged the colon crypts and intestine villi through their outer walls and demonstrated the increase in imaging depth and contrast in biological tissues. Because volumetric gating is the addition of ultrasound-modulated optical detection to confocal reflectance microscopy for the gating of multiply scattered waves, it can be compared with optical coherence microscopy (OCM), in which coherence gating (or temporal gating) is added to confocal reflectance imaging[4–8]. In the case of OCM, the multiply scattered waves traveling to depths shallower than the object plane, and thus having the same flight time as the ballistic wave, cannot be filtered out. The contribution of these ungated multiply scattered waves tends to increase with the depth, thereby setting the depth limit of the OCM. In contrast, volumetric gating eliminates all the multiply scattered waves traveling at depths shallower than the object plane, making this technique particularly suitable for deep imaging. Volumetric gating can also be beneficial for penetrating a scattering medium with a small anisotropic factor or a short scattering mean free path that generates intense multiply scattered waves from shallow physical depths.

The imaging depth of the volumetric gating is primarily determined by the detector dynamic range. The ultrasound-induced modulation efficiency of a light wave is only ~ 1 %, implying that only a small fraction of the ballistic waves and multiply scattered waves traveling through the ultrasound focus contribute to the interferometric signal. The unmodulated contribution, along with the multiply scattered waves and stray reflections traveling outside the ultrasound focus, mostly uses up the detector dynamic range. Although we designed the experimental set-up to block the stray reflections physically using polarization optics and anti-reflection coating in the sample tray and to focus the ballistic wave to a few pixels in the camera by using the confocal detection configuration, this innate limitation is a bottleneck that needs to be resolved. For example, in the measurement of the PSF for the depth of 12.1 $l_s$ as shown in Fig. 4f, the ultrasound-modulated light is only 0.012 % of the total recorded signal, suggesting that almost all the detector dynamic range (1:37,500) was filled with the unmodulated signal. Multiple measurements can increase the effective dynamic range but at the expense of imaging speed. A recent approach based on the use of unmodulated light can be a possible solution to this drawback[22].

Volumetric gating adopts and implements the unique advantage of ultrasound modulation—the gating action occurs within a scattering medium. In other words, it provides a physical probe for sampling light waves reaching a small volume set by the ultrasound focus. In our study, we could map the distribution of the light trajectories within the 3D scattering medium by combining the volumetric gating in reflection mode and lateral gating in transmission mode. The two measurements enabled us to determine the axial gating efficiency uniquely, which provides us the information on the fraction of light waves that penetrate to depths shallower than the ultrasound focus, among all the backscattered waves. This presents the first experimental measurement of the traveling history of light waves inside a 3D scattering medium; such light trajectories were previously accessible only through numerical simulations[17]. Thus, our experimental system and analysis framework are anticipated to serve as important tools for investigating wave propagation in a scattering medium. Volumetric gating is a powerful modality by itself for deep optical imaging, and when combined with temporal/coherence gating, both the gating methods can complement each other, making it possible to reach the ultimate depth limit that can be achieved by using any multiple-scattering gating approach.

## Materials and methods

**Measurement procedure of confocal imaging and image reconstruction**

The objective lens illuminated a beam focused onto the target plane, and the reflected light was collected by the same objective lens and delivered to the sCMOS plane through an optical setup, including 4f optical systems. The complex field of light was then measured using four-step phase shifting interferometry. As each pixel on the target plane and the sCMOS plane had a 1:1 relation, we measured the signal only at the sCMOS pixels conjugate to the focused illumination, corresponding to an optical diffraction-limited image. This procedure is equivalent to the conventional confocal measurement, in which a physical pinhole is used. We tilted the galvano mirrors to implement 2D raster scanning, and finally recorded the 2D image of the target plane.

**Preparation of phantom tissue and sample tray**

To fabricate the scattering media with different scattering mean free paths, we varied the concentration of fat emulsion (Intralipid, Sigma Aldrich I141) added to the PAA gel slab. We prepared a custom-made sample tray with seven slide glasses ($76 \times 25 \times 1$ mm$^3$) attached together to form a housing with a total depth of 7 mm, and the gel slab was inserted into this housing. To introduce ultrasound and prevent the slab from drying or changing in volume, we sealed the open volume of the tray with two slides, to allow light propagation, and a 10-μm-thick linear low density polyethylene film, to modulate the media using ultrasound. Especially, the glass facing the objective lens was coated with anti-reflective layers to minimise stray reflections.

**Preparation of reflective targets**

For the fabrication of reflective targets, we printed the designed patterns on cover glasses through gold deposition. The cover glasses were then spin-coated with poly methyl methacrylate (PMMA) on top of the sputtered conduction layer of platinum for electron-beam lithography. The coated glasses were then exposed to an electron beam, and the patterns were carved to prepare a mould for the deposition of a gold–titanium alloy. After the deposition, we removed the PMMA coating with acetone, which finalised the printing of the target. The thickness of each gold deposited layer was 80 nm. The target was inserted into the PAA gel at a typical depth of 4 mm.

**Preparation of biological tissue**

We extracted intestine tissue from two- to three-day-old Spraque–Dawley rats, and colon tissue from a three-month-old C57BL/6 mice. Each tissue was quickly excised and fixed for 2–4 h at 4 °C in 4 % paraformaldehyde. After fixation, the tissues were washed with a phosphate-buffered saline solution and then sandwiched in the transparent PAA gel slabs (dimensions: ~20($x$) × 10($y$) × 7($z$) mm$^3$). To set up the imaging geometry for examining the inner structures from the outside of the biological samples, the colon and intestine were placed such that their outer surfaces face the objective lens. All the aforementioned experimental procedures and protocols were in accordance with the guidelines established by the Committee of Animal Research Policy of Korea University.

**Acknowledgements**

This work was supported by the Institute for Basic Science (IBS-R023-D1), and Basic Science Research Program through the National Research Foundation of Korea (NRF) funded by the Ministry of Education (2022R1I1A1A01072789), National Research Foundation of Korea (NRF) grant funded by the Korea government (MSIT) (No. NRF-2021R1A5A1032937, 2021R1C1C1011307, 2019R1A2C2004846), the Korea Agency for Infrastructure Technology Advancement (KAIA) grant funded by the Ministry of Land, Infrastructure and Transport (Grant 22NPSS-C163379-02).

## Author contributions

M.J. and W.C. conceived the initial idea. H.K. developed the theoretical modeling, designed the experiments, and analyzed the experimental data with the help of M.J. and W.C. H.K. prepared the sample and carried out the experiments with the help of J.K. J.-H.H. prepared the mouse colon and small intestine specimen. J.K., J.C. and S.L. fabricated the gold-coated reflective targets. All authors contributed to writing the manuscript. W.C. supervised the project.

## Data availability

All relevant data are available from the authors upon request.

## Conflicts of interest

The researcher claims no conflicts of interest.